# Advanced Frequency Identification Power Metering System for Energy Usage

Rozita Teymourzadeh, Ahmed J.A. Abueida, Kok Wai Chan, Mahmud Iwan S, Vee Hoong Mok
Faculty of Engineering, Technology & Built Environment
UCSI University
Kuala Lumpur, Malaysia
rozita@ucsiuniversity.edu.my

*Abstract*—Energy meter measures the amount of power consumed by electrical loads in residential, industrial and commercial applications. In this project, the focus goes to the implementation of a smart power measurement system to allocate identification for individuals and determine the client's energy usage. The incorporation of two PIC 16F877A microcontrollers and radio-frequency identification (RFID) reader in this research work make the system operation smooth and reliable. This paper presents the development of an intelligent prepaid power metering system enabling power utilities to collect electricity bills from consumers prior to the usage of power. Home owners are able to monitor reliable power consumption data for efficient power management. To conclude, a graphical user interface (GUI) has been designed to be applied for data transmission between the personal computer and RFID smart card which allows the credit to be transferred to the smart card.

## I. Introduction

Recent hikes in home energy costs across many countries has led to the increase in global energy awareness among home owners. The smart prepayment metering system becomes the economical substitution to the conventional electromechanical meters for efficient utilization of energy. While enabling home owners to constantly monitor energy consumption data at a desired time interval, the prepaid power metering system facilitates a flexible and accurate billing scheme for electrical power usages. Along with the integration of an intelligent home automation system [1-2], home owners can adjust their electricity consumption behavior by opting to switch off selected appliances remotely during peak periods. The smart prepaid electricity metering system is capable to detect and deter power theft issues which include tampering and modifying of conventional electromechanical meters resulting in lower reading on the meter. In view of the growth in global energy demand, research and development in the electrical field today has been concentrated on the development of innovative metering technologies.

Research work has been conducted in [3] utilizing Microsoft Visual Basic 6.0 in developing a user interface program to issue smart card ID and increase the credit in the customers' card used for the prepaid electricity metering system. ZigBee and GPRS technologies have been proposed in [4] for the design of the wireless remote household metering system. Dang *et al.* [5] explored the implementation of the power metering automatic system, while addressing the complexity of the GPRS/CDMA communication channel to ensure data exchange security and stability of prepaid functions. RFID has been integrated into the power meter and outage recording system in [6]. M. Wasi-ur-Rahman *et al.* in [7] proposed a SMS based remote metering system that utilizes GSM networks amongst many other research works [8-11]. Survey results from a user study about the implementation of a digital feedback system along with a prepaid electricity metering system in a Chinese university dormitory in China have been collected and recently published [12].

## II. Description

Top-up circuitry based package plays a significant role in the frequency identification system in this project. The top-up circuitry consists of DC power supply, regulator, RFID writer/reader module, and serial port. In proposed design, the operation management of the circuitry system takes place internally. In the operation of the RFID top-up system, the RFID reader first detects the RFID card signal and determine the validity of the card. The top-up operation can be achieved by the interfacing between the top-up circuitry and user personal computer (PC) via USB RS232 using the graphical user interface (GUI) window.

The RFID reader/writer module operates at TTL logic level whereas the serial communication in PC works on RS232 standard (±25V). Since the RS232 serial communication in a computer is not compatible with the RFID reader/writer module, MAX232 is utilized to enable the duplex communication between them by converting TTL logic levels to RS232 logic levels. The source of the power for the RFID top-up circuitry is from the DC power supply. The DC power supply will supply 7V to 15V DC to the RFID top-up circuitry. The regulator in the RFID top-up circuitry converts the input DC voltage to +5V. This output voltage will be used as the power supply for MAX232 and the RFID reader/writer device. Fig. 1 shows the fundamental system operation of the RFID based top-up circuitry at a glance.

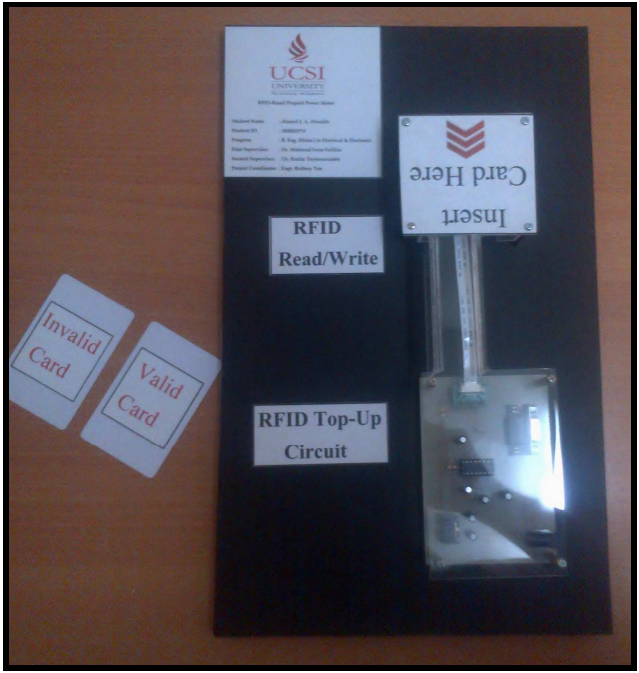

Figure 1. RFID top-up circuitry.

The RFID top-up circuitry is incorporated in the proposed intelligent prepaid power metering system, as illustrated in Fig. 2. Two 8-bit PIC16F877A microcontrollers have been chosen to be integrated with the prepaid electricity metering system for high efficiency usage. The program stored inside the microcontroller contains the protocol for accessing various hardware peripherals such as LCD, relay, GSM module, and etc. The microcontroller is essential in determining the credit balance from the communication between the RFID card and RFID reader, triggering the relay to turn ON the loads, and displaying the credit balance along with consumed power in the LCD display. Another PIC microcontroller is embedded in the prototype of the proposed RFID prepaid power meter which is responsible to send 5V to trigger the GSM modem in sending Short Message System (SMS) text messages as well as produce alert sound from the buzzer when the credit is low. The text messages sent to mobile phones contain feedback status indicating low credit for the prepaid power meter. Failing to transfer credits into the RFID card by consumers upon the credit reaching zero will lead to relays cutting off the supply of electricity. Once the home owners have increased credits in the RFID card using the developed credit top-up graphical user interface, the relay connects home automation and resumes the electrical supply.

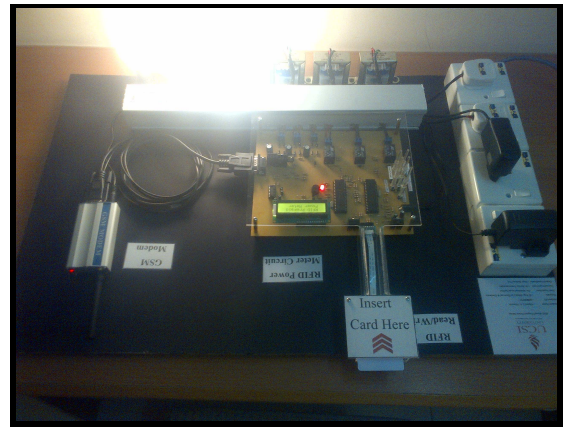

Figure 2. Prototype of the proposed RFID prepaid power meter.

III. STAGE REALIZATION AND SIMULATION

The RFID prepaid power metering algorithm is shown in Fig. 3. The program starts executing with the configuration of two I/O ports for the purpose of reading and writing data. The I/O includes the RFID reader connection to the microcontroller and LCD display.

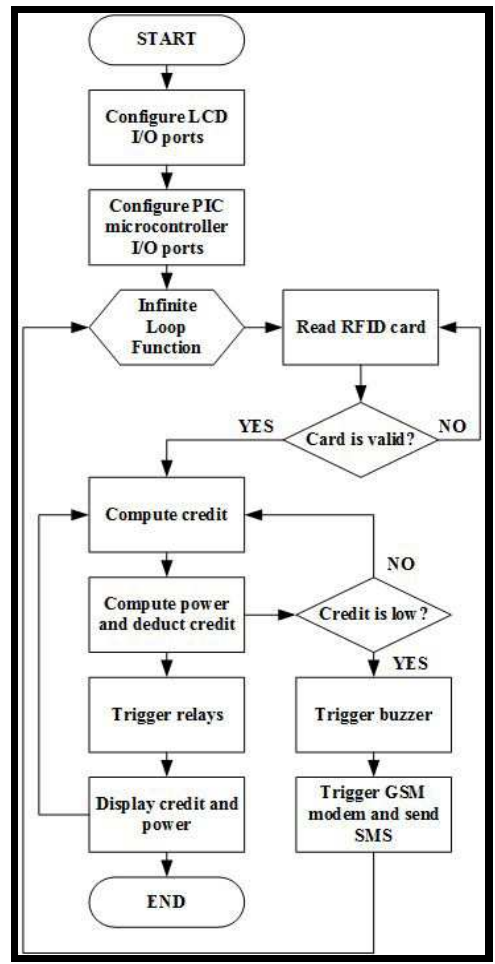

Figure 3. RFID prepaid metering algorithm flowchart.

Once both the I/O ports and the direction of data flow have been determined, the RFID card is read from the RFID reader. The RFID reader transmits an encoded radio signal to interrogate and validate the card. With the validated RFID card, the PIC microcontroller then computes the credit deduction and displays both power usage and credit. Due to the uniqueness of the RFID smart card, the credits will not be transferred to the card if a different RFID card is used. The algorithm is then restarted when the power is turned off.

Fig. 4 demonstrates the simulation of the proposed RFID based prepaid power metering system being implemented using Proteus. A 16x2 character LCD is used to display the remaining credit and total power consumed in this proposed prepaid power meter for the ease of monitoring by home owners. Both RFID reader/writer module and RFID card are not simulated with the proposed RFID based prepaid power metering system. Hence, the value of credit balance and consumed power displayed on the LCD display during simulation are respectively zero.

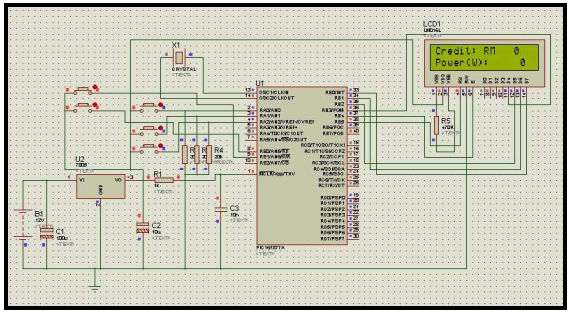

Figure 4. Simulation of the proposed RFID based prepaid metering system.

A feedback path is connected at the load as a current sensing circuit using resistive shunt. Commonly applied in the RFID power meter, the resistive shunt is simply a resistor placed in series with the load. A voltage is developed across the shunt that is directly proportional to the current flowing through the load. A capacitor connected to the output of the step-down transformer converts the alternating current (AC) voltage to the direct current (DC) voltage as illustrated in Fig. 5. Ammeter connected in the circuit is used for the purpose of calibration to observe the output $V_{out}$ in relation to the current when calibrated using different AC loads.

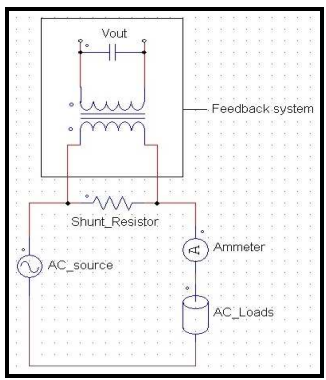

Figure 5. Current sensing feedback path.

The waveform simulation result of the top-up circuitry using Proteus software is illustrated in Fig. 6. A series of pulse generator has been used as an alternative to the RFID writer during simulation to test the operation of the circuit. The simulated waveform resembles an amplitude-shift keying (ASK) signal. The digital signal modulates on the carrier generated by MAX232. The amplitude of the carrier changes with respect to the amplitude of the digital signal. Hence, under magnification, the ASK signal is a sinusoidal waveform.

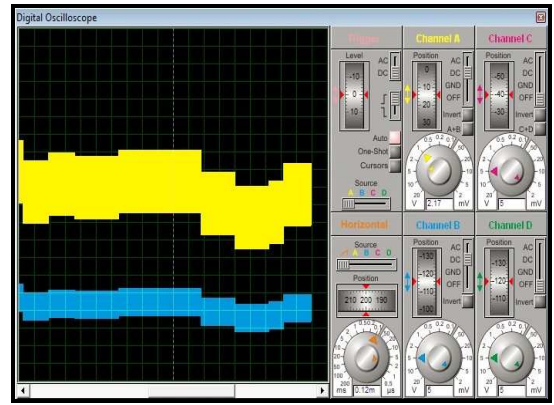

Figure 6. ASK signal generated during simulation of RFID top-up circuitry.

The RFID top-up system is a GUI program developed for the user to increase the credit inside the validated consumer smart card for the prepayment power metering system as shown in Fig. 7. Microsoft Visual Basic 2008 is used as the tool to write the programs for communication between the personal computer and the smart card. COM port 1 has been utilized for the RFID credit top-up program.

In the graphical user interface, it displays the buttons for the user to step-by-step select the actions required to increase the credits on the RFID card. The top-up steps include RFID writer activation, RFID card authentication, reading of the current credit balance on the RFID card, and RFID credit top-up. The current status of the credit top-up process is displayed on the GUI terminal. The RFID interface allows the measured power consumption data to be stored temporarily in the RFID smart card and transferred to the credit top-up graphical user interface.

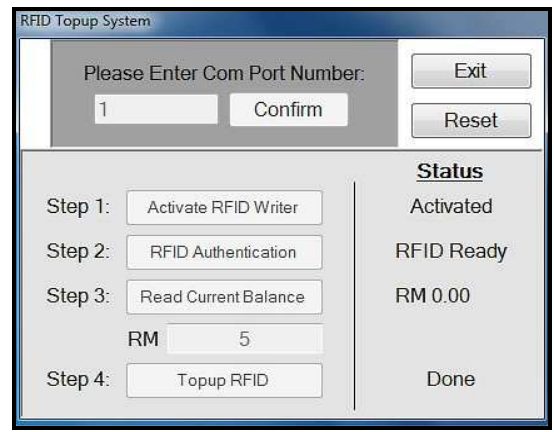

Figure 7. Test run on proposed RFID based prepaid metering system.

With the RFID smart card authenticated, an amount of credit can then be loaded by the user onto the smart card. User is notified about the success of the credit top-up process. The remaining credits inside the smart card after being increased will then be shown on the 16x2 character LCD. Fig. 8 shows that the remaining credit balance on the smart card is RM 5 upon the successful credit top-up process.

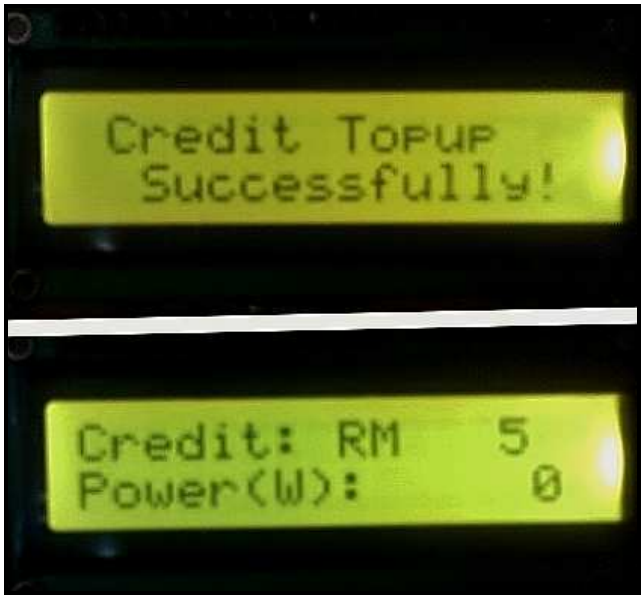

Figure 8. Credit top-up for RFID prepaid meter.

IV. RESULTS AND DISCUSSIONS

Table 1. Experimental power consumption test on RFID power meter under different light bulbs.

| Time (s) | Power Consumption (in Watts) | | |
|---|---|---|---|
| | Light bulb 1 | Light bulb 2 | Light bulb 3 |
| 5 | 57 | 24 | 14 |
| 10 | 57 | 24 | 14 |
| 15 | 57 | 24 | 14 |
| 20 | 57 | 24 | 14 |
| 25 | 57 | 24 | 14 |
| 30 | 57 | 24 | 14 |
| 35 | 0 | 24 | 14 |
| 40 | 0 | 24 | 14 |
| 45 | 0 | 0 | 14 |
| 50 | 0 | 0 | 0 |
| 55 | 0 | 0 | 0 |
| 60 | 0 | 0 | 0 |

Experiment tests were conducted on the operations of the RFID power metering system using different light bulbs and the results were tabled in Table 1. From the experimental test results, it is seen that within 1 minute, the light bulb rated 60 Watts has turned off. This is due to the higher power consumption by the 60W light bulb as compared to the power consumption by 25W and 15W light bulbs respectively. From this experiment test, it can be deduced that higher power consumption will lead to faster credit deduction from the power meter.

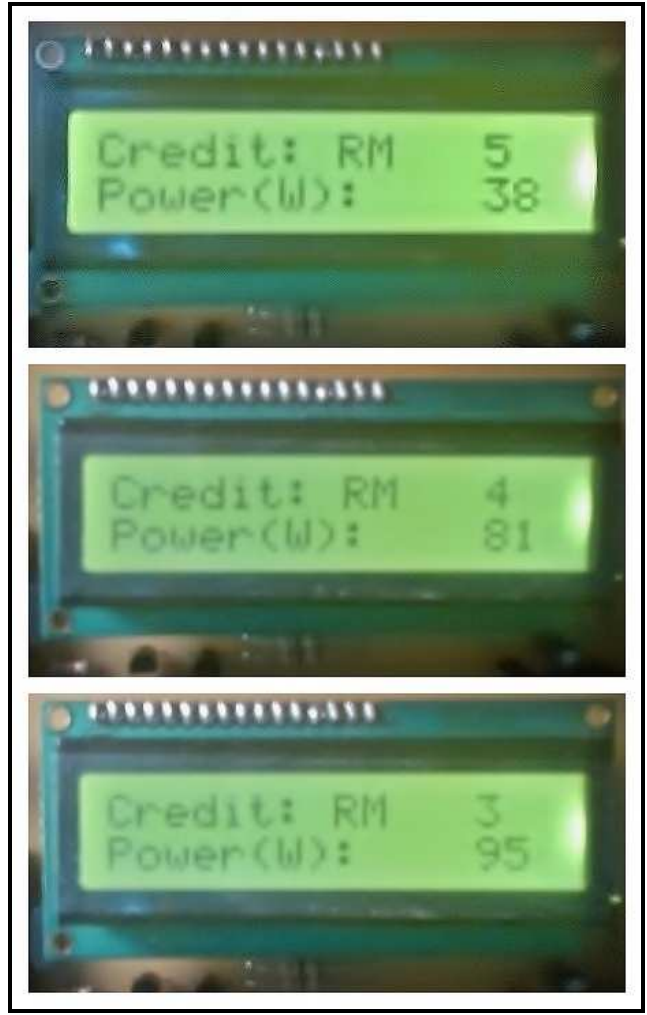

Figure 9. Deduction of credits for RFID prepaid meter.

The experiment was conducted on the proposed power metering system with the credit balance in Fig. 9. As the remaining credit continues to be deducted, total power consumption by electrical loads is displayed from the LCD display for the ease of energy monitoring whilst the remaining credit drops to RM 1 (Fig. 10). The PIC microcontroller then sends 5V to trigger the GSM modem which then sends text messages to home owners indicating that the credit for the power meter system is running low (Fig. 11). The delay of receiving the text message, as had been tested practically [1], does not exceed 2 or 3 seconds due to GSM communication that is mostly governed by the SMS protocol. Concurrently, the alert sound from the buzzer will also be triggered to ensure that the users are aware that the electrical power will be cut off if credit has yet to be increased.

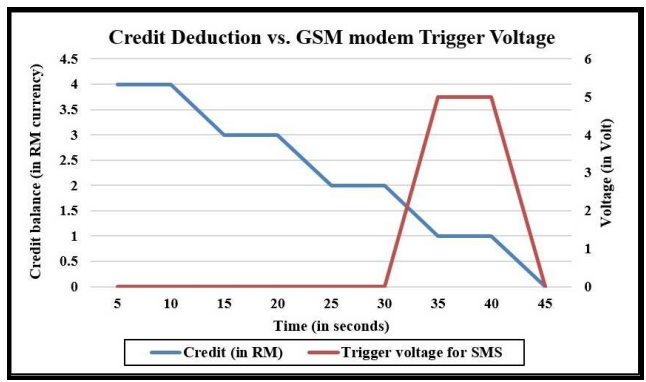

Figure 10. Power metering system behaviour during low credit.

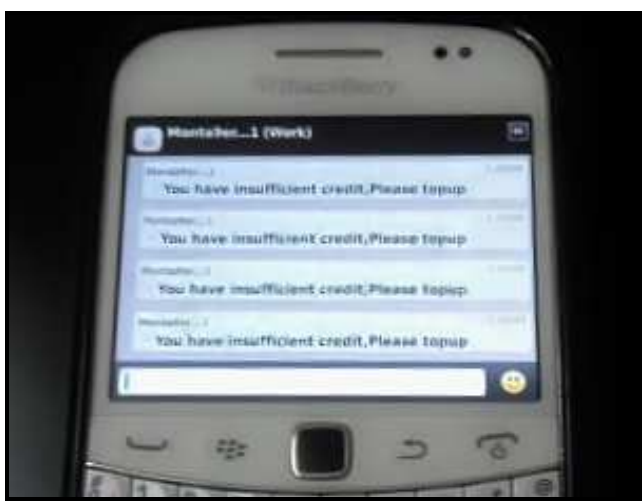

Figure 11. SMS notification of low credit for RFID prepaid meter.

V. CONCLUSION

A RFID based power metering system which includes the credit top-up circuitry has been designed and implemented. The proposed power metering system enables consumers to monitor power consumption data at a desired time interval. With accurate power consumption data, home owners will be able to gain control on daily home energy usage. Experimental tests conducted on the power metering system has shown that the developed graphical user interface ensures consumers pay for their utility bill in a timely manner to prevent the electricity bill to prevent the electrical supply from being cut. With the security uniqueness of the RFID card, this proposed advanced frequency power metering system will be able to deter power theft issues.

REFERENCES

[1] R. Teymourzadeh, S. A. Ahmed, C. Kok Wai, and H. Mok Vee, "Smart GSM based Home Automation System," in *Proc. 2013 IEEE Conference on Systems, Process & Control (ICSPC)*, Kuala Lumpur, Malaysia, pp. 306-309.
[2] S. Ahmad, "Smart metering and home automation solutions for the next decade," in *Proc. 2011 International Conference on Emerging Trends in Networks and Computer Communications (ETNCC)*, Udaipur, pp. 200-204.
[3] B. H. Kwan and M. Moghavvemi, "PIC based smart card prepayment system," in *Proc. 2002 Student Conference on Research and Development (SCOReD 2002)*, Shah Alam, Malaysia, pp. 440-443.
[4] Q. –x. Li and G. Li, "Design of remote automatic meter reading system based on ZigBee and GPRS," in *Proc. of the Third International Symposium on Computer Science and Computational Technology (ISCSCT '10)*, Jiaozuo, China, pp. 186-189.
[5] S. –l. Dang, J. –f. Yang, F. –s. Wei, and H. –k. Que, "Design of prepayment implementation and its data exchange security protection based on Power Metering Automatic system," in *Proc. 2012 IEEE International Conference on Computer Science and Automation Engineering (CSAE)*, Zhangjiajie, pp. 331-335.
[6] C. Shun-Yu, L. Shang-Wen, T. Jen-Hao, and T. Ming-Chang, "Design and implementation of a RFID-based power meter and outage recording system," in *Proc. 2008 IEEE International Conference on Sustainable Energy Technologies (ICSET 2008)*, Singapore, pp. 750-754.
[7] M. Wasi-ur-Rahman, M. T. Rahman, T. H. Khan, and S. M. L. Kabir, "Design of an intelligent SMS based remote metering system," in *Proc. 2009 International Conference on Information and Automation (ICIA'09)*, Zhuhai, Macau, pp. 1040-1043.
[8] G. Bo, L. Xinyu, and X. Junming, "The Design and Realization of Acquisition Equipment in Remote Reading System Base on Low Voltage Power Line Carrier," in *Proc. 2010 International Symposium on Computational Intelligence and Design (ISCID)*, Hangzhou, pp. 24-27.
[9] K. Li, J. Liu, C. Yue, and M. Zhang, "Remote power management and meter-reading system based on ARM microprocessor," in *Proc. 2008 Conference on Precision Electromagnetic Measurements Digest (CPEM 2008)*, Broomfield, Colorado, pp. 216-217.
[10] Z. –h. Zheng, X. –j. Zhou, and W. Zhang, "Design and Implementation of Remote Meter Reading System," in *Proc. 2010 2nd International Conference on Software Technology and Engineering (ICSTE)*, San Juan, Puerto Rico, Volume 2, pp. 247-250.
[11] Q. Zongming, Y. Xiaoshan, and S. Zhenhai, "Database design of GPRS-oriented remote meter reading system," in *Proc. of 2011 International Conference on Computational Problem-Solving (ICCP)*, Chengdu, China, pp. 493-495.
[12] T. Liu, X. Ding, S. Lindtner, T. Lu, and N. Gu, "The Collective Infrastructural Work of Electricity: Exploring Feedback in a Prepaid University Dorm in China," in *Proc. 2013 ACM International Joint Conference on Pervasive and Ubiquitous Computing (UbiComp'13)*, Zurich, Switzerland, pp. 295-304.